\documentclass{revtex4}
\usepackage{graphicx}
\usepackage{fancyhdr}
\usepackage{amsmath}
\usepackage{enumerate}
\usepackage{color}

\begin{document}

\title{Partial Mass Degenerated Model and Spontaneous CP Violation in the Leptonic Sector
}


\author{Hiroyuki Ishida}
\affiliation{Maskawa Institute, Kyoto Sangyo University, Motoyama, Kamigamo, Kyoto 603-8555, JAPAN
\footnote{Current affiliation and address : Tohoku University\,, h\_ishida@tuhep.phys.tohoku.ac.jp\,.}}

\begin{abstract}
We have investigated a flavour model \cite{Araki:2012hb} which inspired by small squared-mass difference measured in solar neutrino oscillation experiments
and observability in neutrinoless double beta decay experiments.
In our model, the $1^{\rm st.}$ and $2^{\rm nd.}$ generations of fermions have a common mass at the leading order.
Such limit may be a good starting point from the points of view of understanding the mixing patterns and mass spectra.
In this limit, the mass matrices are respected an $O (2)$ symmetry on flavor space of the first two generations.
For simplicity, we propose a model for lepton sector based on the $D_N$ group which is a discrete subgroup of $O (2)$.
We show that our model can reproduce the experimental data without hierarchical couplings except for $5 \mathchar`- 10 \%$ tuning partially for the large neutrino mixing.
Further, we show a novel relation between the tiny electron mass and the relatively large $\theta_{13}^{\rm PMNS}$ via CP violation by the complex vacuum expectation values of the extra scalar fields.
\end{abstract}

\maketitle

\thispagestyle{fancy}


%%%%%%%%%%%%%%%%%%%%%%%%%%%%%%%%%%
\section{Introduction and Motivation}
On $4^{\rm th.}$ July 2013, we have been reported Higgs-like boson at long last.
Although we don't know that whether this new boson is the Higgs boson or not, 
we would have found the last piece of the standard model (SM).

Although the SM of particle physics can predict various experiments very accuracy, 
there are some problems and unsatisfactory points which have not been solved yet.
In this paper, we would like to focus on the origin of the flavour structure of fermions among these problems.
When we consider to solve this problem, there is no guiding principle at the present.
For a long while, many scientists have employed discrete flavour symmetries (for a review see ref.\cite{Ishimori:2010au}) in the neutrino sector 
and tryed to understand  the origin by using the geometric structure.
Such symmetries derive tri-bi maximal type mixing \cite{Harrison:2002er} 
and this mixing have been compatible with the neutrino oscillation data before growth of the reactor experiments \cite{rct}.
These experiments have reported non-zero reactor angle, $\theta_{13}^{\rm PMNS}$, at very high confidence level.
At this point, we have to reconsider the origin of the flavour structure of fermions.

The symmetry like flavour symmetry should be broken at enough high energy compared with electroweak scale \cite{no-go}.
Searching for the clue of the guidepost for the flavour structure, 
it is very important to seek a small parameter in the model.
From this point of view, we show the following candidates as small parameters for examples :
\begin{enumerate}
\item[(i)] $\theta_{13}^{\rm PMNS}\ll\theta_{12}^{\rm PMNS}, ~\theta_{23}^{\rm PMNS}$ ~~~({\rm or}~~$|V^{\rm PMNS}_{e3}|\ll {\rm the~others}$),
\vspace{2mm}
\item[(ii)] $|\theta_{23}^{\rm PMNS}-45^\circ|\ll \theta_{23}^{\rm PMNS}$~~~({\rm or}~~$|V^{\rm PMNS}_{\mu 3}-1/\sqrt{2}|\ll |V^{\rm PMNS}_{\mu 3}|$),
\vspace{2mm}
\item[(iii)] $\Delta m^2_{12}=(m_2^\nu)^2 - (m_1^\nu)^2 \ll \Delta m^2_{23} =|(m_3^\nu)^2 - (m_2^\nu)^2|$,
\vspace{2mm}
\item[(iv)] $m_1^{u,d,\ell},~m_2^{u,d,\ell} \ll m_3^{u,d,\ell}$,
\vspace{2mm}
\item[(v)] $\theta_{ij}^{\rm CKM} \ll \theta_{ij}^{\rm PMNS}$~~~({\rm or}~~$|V^{\rm CKM}_{ij}| \ll |V^{\rm PMNS}_{ij}|$),
\end{enumerate}
where $\theta_{ij}^{\rm PMNS}$ and $\theta_{ij}^{\rm CKM}$ stand for the mixing angles in the Pontecorvo-Maki-Nakagawa-Sakata (PMNS), $V^{\rm PMNS}$, and Cabibbo-Kobayashi-Maskawa (CKM), $V^{\rm CKM}$, 
mixing matrices which are the standard parametrization defined by the Particle Data Group \cite{PDG}, 
respectively, and $m_{i}^{f}$ with $f=u,d,\ell,\nu$ and $i=1,2,3$ denote masses of the up-type quarks, down-type quarks, charged leptons and neutrinos.
In this paper, we pay our attentions from (iii) to (v) and discuss the gifts of these viewpoints.

In the limit of $\Delta m_{12}^2 \to 0$, that is $m_1^\nu = m_2^\nu$, 
the $O(2)$ flavour symmetry appears in the Majorana neutrino mass matrix as
\begin{equation}
\begin{split}
\begin{pmatrix}
\cos \theta & \sin \theta &0\\
-\sin \theta & \cos \theta &0\\
0 &0 &0
\end{pmatrix}
\begin{pmatrix}
m_D\\
& m_D &\\
& &m_3
\end{pmatrix}
\begin{pmatrix}
\cos \theta & -\sin \theta &0\\
\sin \theta & \cos \theta &0\\
0 &0 &0
\end{pmatrix}
=
\begin{pmatrix}
m_D\\
& m_D &\\
& &m_3
\end{pmatrix}
\,,
\end{split}\label{eq:O(2)}
\end{equation}
If we do not necessary to make unnatural hierarchy between the masses for first two neutrinos and the third one, it can be realized to obtain the degenerated neutrino masses.
\begin{figure}[t]
\begin{center}
\includegraphics*[width=0.5\textwidth]{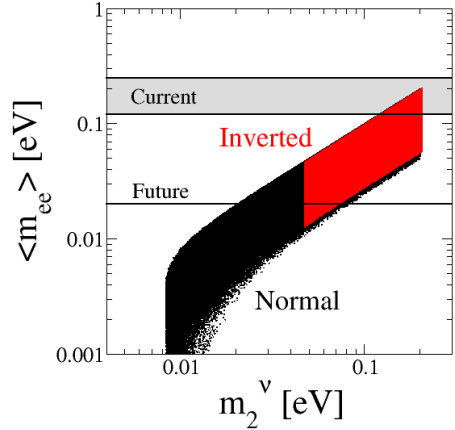}
\end{center}
\vspace{-0.8cm}
\caption{The allowed region of the effective mass, $\langle m_{ee} \rangle$, of $0 \nu \beta \beta$ decay as a function of $m_2^\nu$ in the standard $3\nu$ framework, where 3$\sigma$ constraints of the neutrino oscillation parameters from \cite{fogli} are imposed while varying CP phases from $0$ to $2\pi$.
The current upper bound by combining results from the EXO \cite{exo} and KamLAND-Zen \cite{kam} experiments at $90\%$ C.L. and the expected future bound by the next generation EXO and KamLAND-Zen experiments \cite{neut12} are also shown.
} \label{fig:0nbb}
\end{figure}
Such region of neutrino masses (roughly speaking $m_2^\nu = 0.05{\rm -}0.10 {\rm eV} $) 
is very attractive from the neutrinoless double beta ($0 \nu \beta \beta$) decay.
The effective mass of $0 \nu \beta \beta$ decay can be constrained on the promising region for the next generation experiments for example KamLAND-Zen and EXO experiments shown as FIG.\ref{fig:0nbb}.

On the other hand,  the mass matrix shown as eq.(\ref{eq:O(2)}) seems to be difficult to apply to the charged fermion sector because their masses are strongly hierarchical.
But one can assume the degenerate mass set to be zero at the leading order, 
that is, assign the different doublet representation, ${\bf 2}_{m \neq n}$, or the singlet representation to the right-handed fields.
Such assignments lead two types of mass matrix,
\begin{equation}
M^\ell = 
\begin{pmatrix}
0 &0 &0\\
0 &0 &0\\
0 &0 &m_3^f
\end{pmatrix}
\,\,\,
{\rm or}
\,\,\,
M^\ell = 
\begin{pmatrix}
0 &0 &0\\
0 &0 &0\\
m_{31}^f &m_{32}^f &m_{33}^f
\end{pmatrix}\,.\label{eq:charged}
\end{equation}
We can obtain hierarchical mass for third generation in both case.
In this sense, the partial degenerate limit might be applicable to not only the charged lepton sector but also the quark sectors.
We can connect the small charged fermion masses with the $O(2)$ symmetry breaking.
Also, we may be able to naturally understand some phenomenological relations among the CKM matrix elements \cite{xing}.

The O(2) symmetry breaking triggers flavor mixing at the same time. 
In the quark and charged lepton sectors, 
small flavour mixings are expected because the breaking terms are supposed to be much smaller than the leading terms of eq.(\ref{eq:charged}), 
and thus the observed small CKM mixing could be obtained.
We notice that mixing between the first and second generations is not necessarily small because of the mass degeneracy. This could explain reason why $\theta_{12}^{\rm CKM}$ is a little larger than the others. 
Meanwhile in the neutrino sector, 
flavor mixing can be large since the leading-order neutrino mass matrix in eq.(\ref{eq:O(2)}) is almost proportional to the unit matrix if the neutrino mass hierarchy is mild.
Then, it may be possible to realize the large PMNS mixing even thought the breaking terms are suppressed.

%%%%%%%%%%%%%%%%%%%%%%%%%%%%%%%%%%
\section{Model}
\begin{table}[t]
\begin{center}
{\begin{tabular}{|c|c|c|c|c|c|}\hline
 & $L_I$ & $L_3$ & $\ell_{i}$ & $H$ & $S_I$ \\ \hline
$D_N$ & ${\bf 2}_2$ & ${\bf 1}$ & ${\bf 1}$ & ${\bf 1}$ & ${\bf 2}_1$ \\ \hline
\end{tabular}}
\caption{Particle contents and charge assignments of the model, where $I=1,2$ and $i=1\cdots 3$ denote the indices of generations, $L$ and $\ell$ represent the left- and right-handed SM leptons, and $H$ and $S_{1,2}$ are the SM Higgs and gauge singlet scalar fields, respectively.}
\label{tab:1}
\end{center}
\end{table}
In this section, we briefly explain about our model.
At the beginning, we summarize the charge assignments of each fields in TABLE.{\ref{tab:1}}. 
We adopt the second type of the mass matrix written in eq.(\ref{eq:charged}) for the charged lepton by assigning the singlet representation into the right-handed charged leptons, $\ell_i$, in this analysis.
Our model is constructed with respect to $D_N$ flavour symmetry which is discrete subgroup of $O(2)$.
$D_N$ and $O(2)$ have same multiplication rules.
We introduce a new SM gauge singlet scalar fields, $S_I$, to break this flavour symmetry.
Furthermore, we impose this new scalar fields have complex vacuum expectation values (VEVs) and denote the CP violating phases as $\phi_{1,2}$.
The Lagrangian under the $D_N$ flavour symmetry is given by
\begin{equation}
\begin{split}
{\cal L}_f 
&= 
  y_{i}^{0} ~\overline{L}_3^{} H \ell_i^{}
+ \frac{y^{}_{i}}{\Lambda_F^2} \overline{L}_I^{} H \ell_{i}^{} (S_{}^2)_I
+ \frac{y^{'}_{i}}{\Lambda_F^2} \overline{L}_I^{} H \ell_{i}^{} (S_{}^{*2})_I
+ \frac{y^{''}_{i}}{\Lambda_F^2} \overline{L}_I^{} H \ell_{i}^{} (|S_{}|^2)_I \\
& 
+ \frac{f_\nu}{\Lambda_\nu}L_I L_I H H
+ \frac{f_\nu^\prime}{\Lambda_\nu}L_3 L_3 H H \\
&
+ \frac{g_\nu}{\Lambda_\nu \Lambda_F^2}L_3 L_I H H (S^2)_I
+ \frac{g_\nu^\prime}{\Lambda_\nu \Lambda_F^2}L_3 L_I H H (S^{*2})_I
+ \frac{g_\nu^{''}}{\Lambda_\nu \Lambda_F^2}L_3 L_I H H (|S|^2)_I \\
&
+ \frac{h_\nu}{\Lambda_\nu \Lambda_F^4}(L_J L_K)_I H H (S^4)_I
+ \frac{h_\nu^\prime}{\Lambda_\nu \Lambda_F^4}(L_J L_K)_I H H (S^{*4})_I
+ \frac{h_\nu^{''}}{\Lambda_\nu \Lambda_F^4}(L_J L_K)_I H H (|S|^4)_I \\
&
+ \frac{h_\nu^{'''}}{\Lambda_\nu \Lambda_F^4}(L_J L_K)_I H H (S^2|S|^2)_I
+ \frac{h_\nu^{''''}}{\Lambda_\nu \Lambda_F^4}(L_J L_K)_I H H (S^{*2}|S|^2)_I \,,
\end{split}
\end{equation}
where $y_i$, $f_\nu$, $g_\nu$ and $h_\nu$ are dimensionless couplings, 
$\Lambda_F$ describes a breaking scale of the $D_N$ flavor symmetry and we have ignored the next-to-next-to-leading terms in the charged lepton sector.
We note that due to an extra charge of $D_N$ which is described as a subscript in TABLE.{\ref{tab:1}}, 
we can construct ideal terms with Froggatt-Nielsen \cite{fn} like mechanism.
Furthermore, we adopt the dimension five Weinberg operator with an energy scale with a scale, $\Lambda_\nu$, for the neutrino masses.

For concluding of explanation of our model, we have to note that our model do not need to have any hierarchical couplings to reproduce the experimental data.
$5 \mathchar`- 10 \%$ tuning between $f_\nu$ and $f_\nu'$, however, 
is necessary in order to realize the large neutrino mixings.
We have checked that our model can reproduce the experimental data by numerical analysis 
with following parameter spaces : 
\begin{eqnarray}
&&y^0_3 = 1.0,~y^0_1 = y^0_2 = 1.2,~
y_1 = -y_2 = y_3  = y_1^\prime = y_2^\prime = -y_3^\prime = 0.8,
\nonumber \\
&&y_1^{''} = -y_2^{''} = -y_3^{''} = 0.8\sim 1.3,
\\
&&f_\nu^\prime=1.0,~f_\nu=0.90\sim 0.95,~
g_\nu = g_\nu^{''} =0.9,~g_\nu^\prime = 0.8\sim 1.3,
\nonumber \\
&&h_\nu = h_\nu^\prime = h_\nu^{''} = -h_\nu^{'''} = -h_\nu^{''''} = 0.8 \sim 1.3,
\\
&&\frac{s_{1,2}}{\Lambda_F} = 0.15\sim 0.30,~
\phi_{1,2}= 0\sim 2\pi ,
\end{eqnarray}
for the case of normal neutrino mass ordering and
\begin{eqnarray}
&&y^0_3 = 1.0,~y^0_1 = y^0_2 = 1.2,~
y_1 = -y_2 = y_3  = y_1^\prime = y_2^\prime = -y_3^\prime = 0.8,
\nonumber \\
&&y_1^{''} = -y_2^{''} = -y_3^{''} = 0.8\sim 1.3,
\\
&&f_\nu^\prime=1.0,~f_\nu=1.05\sim 1.10,~
g_\nu = g_\nu^{''} =-0.9,~g_\nu^\prime = 1.0 \sim 1.5,
\nonumber \\
&&h_\nu = h_\nu^\prime = h_\nu^{''} = -h_\nu^{'''} = -h_\nu^{''''} = -(1.0 \sim 1.5),
\\
&&\frac{s_{1,2}}{\Lambda_F} = 0.15\sim 0.30,~
\phi_{1,2}= 0\sim 2\pi ,
\end{eqnarray}
for the inverted hierarchy case.
\begin{figure}[t]
\begin{center}
%\hspace{-0.5cm}
\includegraphics[width=6cm,clip,angle=-90]{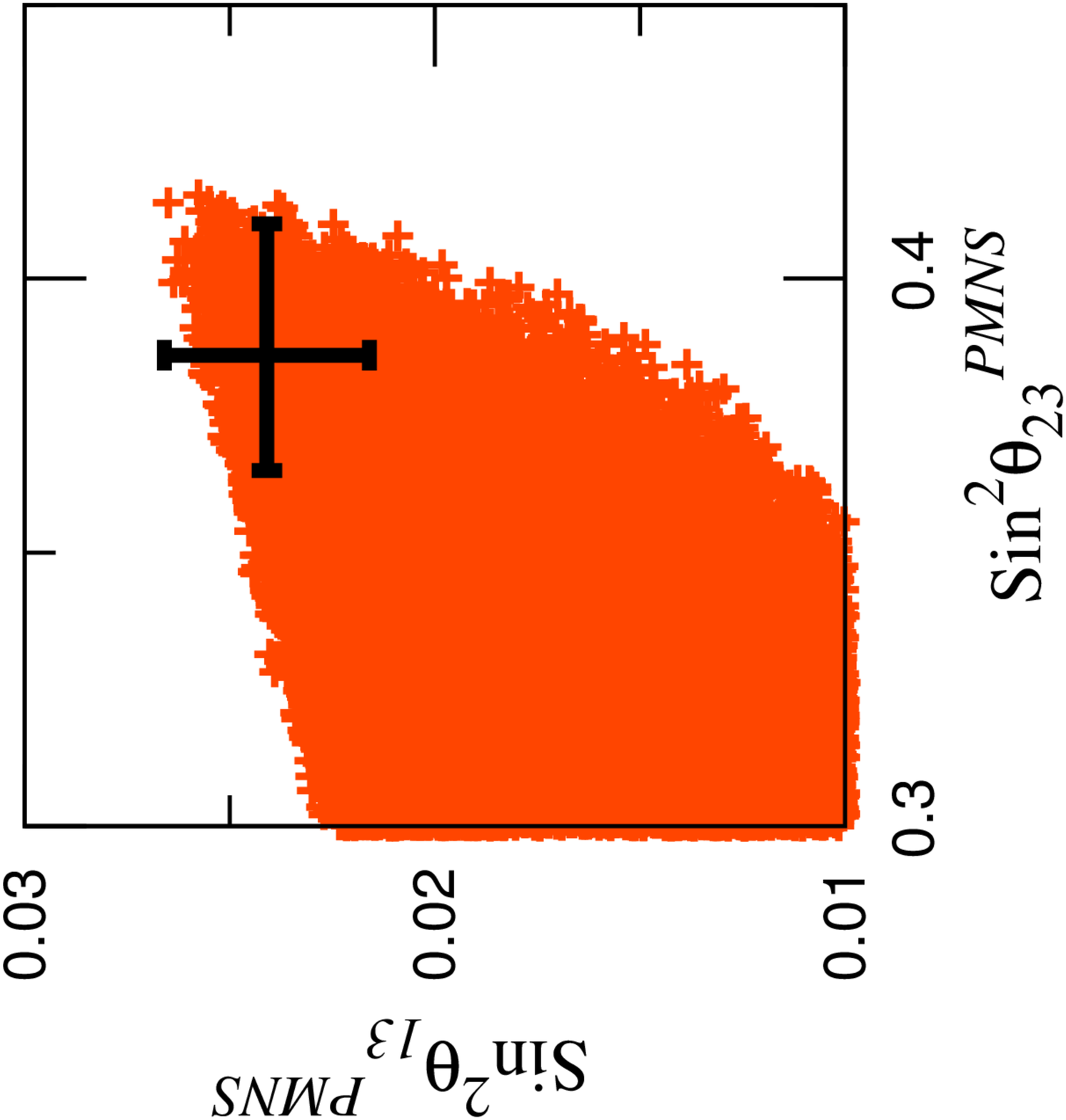}
\hspace{-2cm}
\includegraphics[width=6cm,clip,angle=-90]{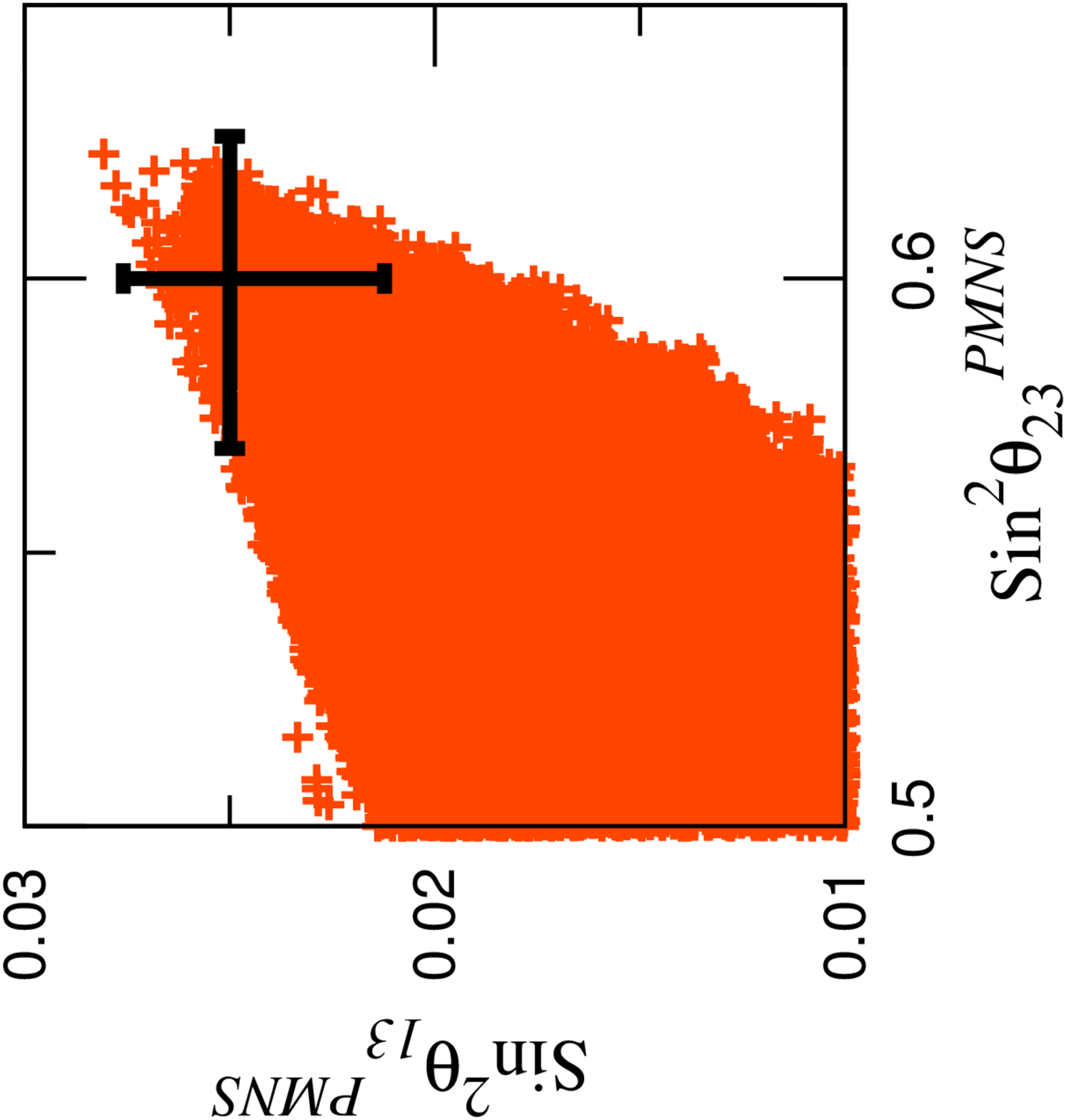}
\vspace{-0.3cm}
\caption{$\sin^2\theta_{13}^{\rm PMNS}$ as a function of $\sin^2\theta_{23}^{\rm PMNS}$ for the cases of normal (left-panel) and inverted (right-panel) neutrino mass orderings.
The 1$\sigma$ error bars are also plotted for the normal and inverted cases from ref.\cite{fogli} and ref.\cite{valle}, respectively.
} \label{fig:rct-atm}
\end{center}
\end{figure}
Actually, we have enough number of parameters to reproduce experimental data shown in FIGs.\ref{fig:rct-atm}.
When we determine the specific generation mechanism for the neutrino masses for example introducing the right-handed neutrino or something like that, 
we may reduce the parameters and construct predictive model.

%%%%%%%%%%%%%%%%%%%%%%%%%%%%%%%%%%
\section{Discussion and Conclusion}
We study a $D_N$ flavour symmetric model inspired by the smallness of solar neutrino mass square difference comparing with atmospheric one 
and testability of the $0 \nu \beta \beta$ decay.
In this model, we introduce a new SM gauge singlet scalar with complex VEVs.
This model has a non-trivial feature between electron mass and a mixing angle, $\theta_{13}^{\rm PMNS}$, via CP violating phases in these.
Here, we assume that $\phi_2 = \phi_1 + \delta \phi$ with $\delta \phi \ll 1$ 
and from this assumption we can rewritten the mass matrix of charged lepton as 
\begin{equation}
\begin{split}
%\makebox[-4mm][r]{}
M^\ell \supset
\begin{pmatrix}
Y_1 \left( s_1^2 - s_2^2 \right) &\,Y_2 \left( s_1^2 - s_2^2 \right)  &\,Y_3 \left( s_1^2 - s_2^2 \right) \\
Y_1 2 s_1 s_2 &\,Y_2 2 s_1 s_2 &\,Y_3 2 s_1 s_2\\
0 &\,0 &\,0
\end{pmatrix}
+ 
i \delta \phi 
\begin{pmatrix}
-Y_1^{\prime} 2 s_2^2 &\-Y_2^{\prime} 2 s_2^2 &\,-Y_3^{\prime} 2 s_2^2\\
Y_1^{\prime} 2 s_1 s_2 &\,Y_2^{\prime} 2 s_1 s_2 &\,Y_3^{\prime} 2 s_1 s_2\\
0 &\,0 &\,0
\end{pmatrix}\,,
\end{split}
\end{equation}
where $Y_i \equiv y_i e^{2i\phi_1} + y_i' e^{-2i\phi_1} + y_i^{\prime \prime}$ 
and $Y_i' \equiv y_i e^{2i\phi_1} - y_i' e^{-2i\phi_1}$.
We can see that each term is ${\rm rank} = 1$ matrix.
The first term is mainly responsible for muon mass and the second one is for electron mass, respectively.
We suppose that $\delta \phi \ll 1$ so that we can obtain the small electron mass.
Furthermore, we need one more assumption.
To avoid accidental contribution to $1$-$2$ mixing from charged lepton mass matrix, 
we assume $s_1 \simeq s_2$.

On the other hand, in the basis in which upper-left $2 \times 2$ elements of $M^\ell M^{\ell \dagger}$ is diagonalized the $1$-$3$ element of neutrino mass matrix can similarly be rewritten as 
\begin{equation}
( \mathcal{M} )_{13} 
\propto 
i \delta \phi
\left\{ 
\left[ g_\nu e^{2i \phi_1} 
- 
g_\nu' e^{-2i \phi_1} 
\right] 
2 s_1s_2^3\, 
\cdots 
\right\}\,.
\end{equation}
As we can see from this equation, the angle $\theta_{13}$ is controlled by 
the small CP violating parameter $\delta \phi$.

We can see that our model can reproduce the experimental data with some assumptions 
but we need to check whether such assumptions are valid or not.
To do this, we discuss the possibility of spontaneous CP violation (SCPV).
At first, we assume that the singlet scalar fields were completely decoupled from the theory at high energy.
In this condition, the only to do is to investigate only the potential of the singlet scalar fields.
One can obtain the scalar potential of $S$ as
\begin{equation}
\begin{split}
V_S &= 
\alpha_S ( s_1^2 + s_2^2 ) 
+ \alpha_S' ( s_1^2 \cos 2 \phi_1 + s_2^2 \cos 2 \phi_2 ) 
+ \beta_S^a ( s_1^2 + s_2^2 )^2 \notag\\
& 
- 4 \beta_S^b s_1^2 s_2^2 \sin^2 ( \phi_1 - \phi_2 ) 
+ \beta_S^c \{ ( s_1^2 - s_2^2 )^2 + 4 s_1^2 s_2^2 \cos^2 ( \phi_1 - \phi_2 ) \} \notag\\
& 
+ \beta_S' \left[ s_1^4 \cos 4 \phi_1 + s_2^4 \cos 4 \phi_2 + 2 s_1^2 s_2^2 \cos [ 2 ( \phi_1 + \phi_2 ) ] \right] \notag\\
& 
+ \gamma_S ( s_1^2 + s_2^2 ) ( s_1^2 \cos 2 \phi_1 + s_2^2 \cos 2 \phi_2 ) \,, \\
\end{split}
\end{equation}
where the couplings $\beta_S^A\, (A=a,b,c)$ distinguish different combinations of $S_{1\,,2}$ 
under the $D_N$ tensor product rules and all of the couplings are supposed to be real.
The minimization conditions respect with the phases are given as 
\begin{equation}
\begin{split}
\frac{\partial V_S}{\partial \phi_1} &=
  -2s_1^2\left[~\{ \alpha_S' + \gamma_S \left( s_1^2 + s_2^2 \right) \} \sin 2 \phi_1 
+ 2 ( \beta_S^b + \beta_S^c ) s_2^2 \sin \left[ 2 \left( \phi_1 - \phi_2 \right) \right] \right.\\
&\left. + 2 \beta_S' \left( s_1^2 \sin 4 \phi_1 + s_2^2 \sin \left[ 2 \left( \phi_1 + \phi_2 \right) \right] \right)~\right]=0\,,\\
\frac{\partial V_S}{\partial \phi_2} &= 
  -2s_2^2\left[~\{ \alpha_S' + \gamma_S \left( s_1^2 + s_2^2 \right) \} \sin 2 \phi_2 
- 2 ( \beta_S^b + \beta_S^c ) s_1^2 \sin \left[ 2 \left( \phi_1 - \phi_2 \right) \right] \right] \\
&\left.+ 2 \beta_S' \left( s_2^2 \sin 4 \phi_2 + s_1^2 \sin \left[ 2 \left( \phi_1 + \phi_2 \right) \right] \right)~\right]=0 \,.
\end{split}
\end{equation}
For simplicity, we set $\alpha_S' = \beta_S' = \gamma_S =0$.
From these condition, the minimization conditions turn to be 
\begin{equation}
2(\beta_S^b + \beta_S^c)s_2^2\sin[2(\phi_1 - \phi_2)]=0\,,~~
2(\beta_S^b + \beta_S^c)s_1^2\sin[2(\phi_1 - \phi_2)]=0\,.
\end{equation}
This analysis is very naive one but 
we can see that the assumption $\phi_1 \simeq \phi_2$ is very good perspective from potential analysis.

For summarizing this paper, 
we would like to see our future works at a glance.
In the present work, 
we adopt the $D_N$ flavour symmetry in our model in order to 
focus on only the flavour structure of fermions.
It is, of course, very interesting and challenging that 
to construct the model with $O(2)$ and take into account new gauge bosons.
Actually, there may exist gauge anomalies.
Furthermore, the quark sector should be included.
We have to check that whether the lightest up- and down-type quarks 
also obtain their masses through CP violation.

We do not mention about the specific generation mechanism of neutrino masses in this work.
The neutrinos must have their masses from the experimental facts and its generation mechanism is discussed by much amount of frameworks.
We have to decide it and discuss the implement of solving the problem for example the origin of the baryon asymmetry of the universe.
These issues will be studied elsewhere.

\begin{acknowledgments}
The author thanks the organizers of HPNP2013 for giving the chance to present our poster at this fruitful conference. 
\end{acknowledgments}

\bigskip % extra skip inserted

\end{document}